# Nanocarbon–Based Photovoltaics

**Marco Bernardi,[1,]\* Jessica Lohrman,[2,]\* Priyank V. Kumar,[1] Alec Kirkeminde,[2] Nicola Ferralis,[1] Jeffrey C. Grossman,[1,†] and Shenqiang Ren[2,†]**

[1] Department of Materials Science and Engineering, Massachusetts Institute of Technology, 77 Massachusetts Avenue, Cambridge, MA 02139-4307

[2] Department of Chemistry, University of Kansas, 1251 Wescoe Hall Drive, Lawrence, KS 66045

\*These authors contributed equally to this work.

[†]e-mail: jcg@mit.edu, shenqiang@ku.edu

## ABSTRACT

Carbon materials are excellent candidates for photovoltaic solar cells: they are Earth-abundant, possess high optical absorption, and superior thermal and photostability. Here we report on solar cells with active layers made solely of carbon nanomaterials that present the same advantages of conjugated polymer-based solar cells – namely solution processable, potentially flexible, and chemically tunable – but with significantly increased photostability and the possibility to revert photodegradation. The device active layer composition is optimized using *ab-initio* density functional theory calculations to predict type-II band alignment and Schottky barrier formation. The best device fabricated is composed of $PC_{70}BM$ fullerene, semiconducting single-walled carbon nanotubes and reduced graphene oxide. It achieves a power conversion efficiency of 1.3% – a record for solar cells based on carbon as the active material – and shows significantly improved lifetime than a polymer-based device. We calculate efficiency limits of up to 13% for the devices fabricated in this work, comparable to those predicted for polymer solar cells. There is great promise for improving carbon-based solar cells considering the novelty of this type of device, the superior photostability, and the availability of a large number of carbon materials with yet untapped potential for photovoltaics. Our results indicate a new strategy for efficient carbon-based, solution-processable, thin film, photostable solar cells.

**KEYWORDS:** photovoltaics, carbon, photodegradation, *ab-initio* materials design

The energy generated from solar photovoltaics (PV) amounts to less than 1% of the total worldwide energy usage at present, for the main reason that producing a kWh of energy from PV panels costs significantly more than burning fossil fuels.[1] Despite an impressive learning curve for PV technology as well as recent advances that have brought Si solar cells ever closer to the single–band gap efficiency limit of ~30%,[1–3] grid-parity PV remains an enormous challenge in most parts of the world. As possible alternatives to inorganic semiconductor PV technology, a number of new materials have emerged: *e.g.*, solar cells based on conjugated polymers,[4,5] small molecules[6] and colloidal nanocrystals[7] are justified by the possibility of utilizing thin film (< 1 μm) materials with high optical absorption, as well as using light–weight flexible substrates, printable organic inks, low temperature and ambient pressure fabrication, enabling reduced device and balance of system costs.[2,4] The ability to use chemical vapor or solution deposition processing is particularly exciting since products such as paper, textiles, automobiles, and building materials could be coated with PV devices thus making solar cells ubiquitous.

Carbon, one of the few elements known since antiquity, holds remarkable potential as a material for solar cells.[8] It's abundant in the Earth's crust (~0.2 wt. %),[9] it can be found in Nature in its elemental form as graphite, diamond and coal, and it is widely used technologically with a record production among other elements of 9 Gt/year.[9] Nanostructured carbon allotropes have been intensively investigated in the past two decades, including single–walled carbon nanotubes (SWCNT),[10,11] fullerenes,[12] graphene[13,14] and their chemical derivatives. These materials hold record values for physical properties important for PV such as carrier mobility, thermal conductivity, mechanical strength, and optical absorption, and are appealing for PV as they can be dissolved in organic solvents to deposit thin solar cell active layers from solution. Other carbon allotropes such as amorphous carbon, nanodiamonds and graphene can be deposited in thin film form on flexible substrates using chemical vapor deposition.[8]

While there has been intense focus on the use of carbon nanomaterials in areas such as electronics and photonics,[15] the potential of carbon as the active layer material in PV is still largely unexplored. In PV, carbon materials have been mainly used as acceptors in polymer–based solar cells[4–6,8,16] or as transparent electrodes,[8] and only recently as the main active layer components in polymer–free solar cells.[17–21] In these works, $C_{60}$ or $C_{70}$ fullerenes were evaporated to form bi-layer devices in combination with SWCNT[17–19] or a composite of SWCNT/rGO/fullerene deposited from aqueous solution,[20,21] with the highest reported efficiency to date of 0.21% for $C_{60}$ (ref. 20) and 0.85% for $C_{70}$ (ref. 21) for such polymer–free, carbon–based devices.

# RESULTS

Figure 1a shows a schematic of the carbon–based solar cells fabricated in this work, which achieve a maximum efficiency of 1.3%. The active layer is entirely deposited from solution and is composed of semiconducting SWCNT (s–SWCNT), the fullerene derivative PCBM,[4] and reduced graphene oxide (rGO),[22] forming a bulk–heterojunction morphology[4] made entirely of carbon nanomaterials. The active layer does not contain other constituents such as conjugated polymers or small molecules, and the atomic carbon concentration in the active layer is as high as $80 - 90$ at. %, in contrast with carbon concentrations of $40 - 50$ at. % for a typical solar cell based on P3HT polymer. The carbon material combinations considered here include small diameter s–SWCNT (diameter $d = 0.75$–1.2 nm) or large diameter s–SWCNT ($d > 1.2$ nm), together with either $PC_{60}BM$ or $PC_{70}BM$ (the latter is used for enhanced optical absorption in the visible),[23] and with or without the addition of rGO. Optimized carrier extraction is achieved by using hole transport and electron blocking layers in conjunction with transparent indium tin oxide (ITO) and Al contacts (see Methods).

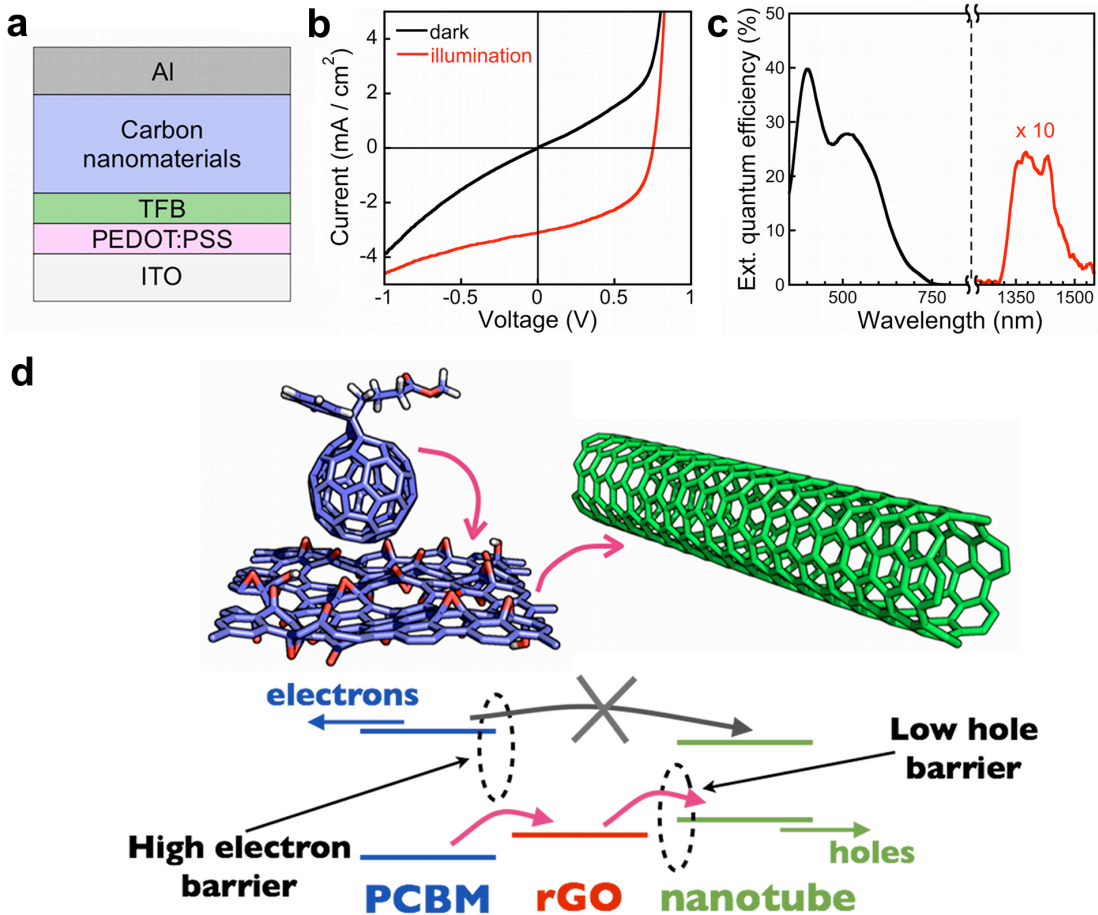

**Figure 1.** (a) Structure of carbon solar cells prepared in this work. Al is the top contact, TFB and PEDOT:PSS are respectively an electron–blocking and a hole–conducting layer deposited on top of the ITO bottom contact. For the best–efficiency device, the active layer, denoted by "carbon nanomaterials", contains a blend of rGO, $PC_{70}BM$, and s–SWCNT with a diameter of 1.2–1.7 nm. (b) Current–voltage curves in the dark and under simulated solar light illumination (1 Sun, AM1.5 spectrum) of the best–efficiency device studied in this work. (c) External quantum efficiency of the best–efficiency device, showing contributions from the $PC_{70}BM$ at visible wavelength (black curve), and from the nanotube $S_{11}$ optical transitions centered at 1400 nm in the infrared (red line, multiplied by 10 for clarity). (d) Interface of PCBM / rGO sheet / s–SWCNT, representing schematically the active layer components of the best–efficiency device. The pink arrows trace the carrier transfer cascade of holes photogenerated within PCBM, selectively injected to rGO due to a large Schottky barrier for electrons, and then transferred to s–SWCNT. The schematic band diagram below shows the same process together with the HOMO and LUMO orbital energies for PCBM and carbon nanotubes, and the Fermi energy or rGO. The crossed gray arrow indicates the absence of electron flow from PCBM to s-SWCNT.

The best–efficiency device has a composition of $PC_{70}BM$ (95 wt. %) / rGO (2 wt.

%) / s–SWCNT (3 wt. %, $d > 1.2$ nm); it achieves an efficiency of 1.3%, deriving from a short–circuit current ($J_{sc}$) of 3.1 mA/cm$^2$, an open circuit voltage ($V_{oc}$) of 0.75 V, and a fill factor (FF) of 0.55 (Figure 1b and Table 1). For comparison, a control sample constituted only by PC$_{70}$BM / rGO (Table 1) without nanotubes showed an efficiency of 0.009%, more than two order of magnitudes lower than our best device. This indicates the key role of s-SWCNT, as hole transport material as well as infrared absorber. The external quantum efficiency (EQE, Figure 1c) shows contributions from both the s–SWCNT donor and the PC$_{70}$BM acceptor. The EQE in the visible derives from optical transitions in the PC$_{70}$BM, with a main peak at ~400 nm and a shoulder peak at 550 nm.[23] The EQE peak in the infrared centered at 1400 nm is due to the $S_{11}$ optical transitions in s–SWCNT with 1.2–1.7 nm diameter, as found in our previous work using the same nanotubes in a polymer–based device.[16] The PV operating mechanism involves a photogenerated hole carrier cascade from PC$_{70}$BM to rGO and then to s–SWCNT (Figure 1d), as explained below. We have fabricated and tested a large set of devices with different ratios and types of PCBM, s–SWCNT and rGO; their performance is summarized in Table 1, and additional EQE curves for some devices (see Table 1) are reported in Supplementary Figure S2.

| COMPOSITION (wt. %) | $J_{sc}$ (mA/cm$^2$) | $V_{oc}$ (V) | FF | EFFICIENCY (%) |
|---|---|---|---|---|
| PC$_{60}$BM (90–99%) / D1–s–SWCNT (1–10%) | 1.2 | 0.6 | 0.24 | 0.17 |
| PC$_{70}$BM (90–99%) / D1–s–SWCNT (1–10%) (*) | 2.2 | 0.55 | 0.35 | 0.42 |
| PC$_{60}$BM (90–99%) / D2–s–SWCNT (1–10%) | 0 | 0 | 0 | 0 |
| PC$_{70}$BM (90–99%) / D2–s–SWCNT (1–10%) | 0 | 0 | 0 | 0 |
| PC$_{70}$BM (98%) / rGO control sample | 0.17 | 0.21 | 0.24 | 0.009 |
| PC$_{60}$BM (88–97%) / D1–s–SWCNT (1–10%) / rGO (~2%) (*) | 0.12 | 0.8 | 0.43 | 0.042 |
| PC$_{60}$BM (88–97%) / D2–s–SWCNT (1–10%) / rGO (~2%) | 1.39 | 0.73 | 0.61 | 0.62 |
| PC$_{70}$BM (88–97%) / D2–s–SWCNT (1–10%) / rGO (~2%) (**) | 3.1 | 0.75 | 0.55 | 1.3 |

**Table 1.** Shown are several tested active layer compositions and the corresponding $J_{SC}$, $V_{OC}$, FF and power conversion efficiency for the best–efficiency device prepared for each material combination. Over 10 devices of each kind were fabricated and tested, yielding a 5% standard deviation on the efficiency. D1–s–SWCNT refers to small diameter nanotubes ($d$ of 0.75–1.2 nm) and D2–s–SWCNT refers to large diameter nanotubes ($d > 1.2$ nm) used in this work. The symbol (*) indicates that EQE curves for these devices are available in Supporting Figure S2. The symbol (**) in the last row refers to the best–efficiency device prepared in this work and shown in Figure 1, achieving an efficiency of 1.3%.

We employed first–principles calculations (Figure 2 and Supplementary Figure S1) to design the optimal combination of carbon nanomaterials used to prepare the device shown in Figure 1. For a bulk heterojunction (BHJ) solar cell containing SWCNT in the active layer, it is crucial to avoid the presence of metallic nanotubes to avoid short–circuiting of the electrodes[16,24] and exciton quenching.[18] For this reason, s–SWCNT are used in this work. However, for a two–component active layer made of s–SWCNT and PCBM, we show that the band alignment (calculated using density functional theory,[25,26]

DFT; see Methods) depends on the nanotube diameter $d$. The HOMO and LUMO level energy offsets ($\Delta E_V$ and $\Delta E_C$ respectively, Figure 2a) vary not only due to the nanotube work function and band gap variation with $d$, but also due to a charge redistribution causing the formation of an interface dipole, thus requiring *ab-initio* calculations of the full interface to compute the band offsets. For a two–phase active layer of s–SWCNT/PCBM, a type–II alignment is ideal for PV operation as it allows the s–SWCNT to work as the donor and the PCBM to work as the acceptor in a BHJ solar cell,[4] thus leading to favorable dissociation of excitons photogenerated in either material. We analyze the case of $PC_{60}BM$ for convenience (due to its higher symmetry, the interfaces with s–SWCNT can be better defined), though $PC_{70}BM$ yields the same trends both in the calculations and in the experiments due to its very similar electronic structure to $PC_{60}BM$.

For all nanotubes studied here ($d$ = 0.75–1.7 nm), the nanotube HOMO level is found to be higher in energy than the HOMO of PCBM (*i.e.* $\Delta E_V > 0$ in Figure 2a), and thus the HOMO orbital of the interface is localized on s–SWCNT. For small $d$ between 0.75–1.2 nm, positive conduction band offsets are found, as confirmed by the interface LUMO orbitals localized on the nanotube (Figure 2a). A type-II alignment favorable for PV operation is thus predicted for nanotube diameters smaller than 1.2 nm. In particular, for the (6,5) nanotube ($d$ = 0.75 nm) constituting up to 50% of the small diameter s-SWCNT sample used in our experiments, the alignment is predicted to be type–II with $\Delta E_C \approx 0.2$ eV, as shown in Fig. 1a; this is consistent with a rigid band alignment model yielding $\Delta E_C = 0.6$ eV for the (6,5) nanotube / $C_{60}$ interface as obtained in ref. 19, given the $0.4 - 0.5$ eV difference in the electron affinity of PCBM and $C_{60}$.[27] We also observe that the small-diameter (10,0) nanotube ($d \approx 0.8$ nm) is an outlier and yields type-I alignment within our DFT calculation. However, even if the (10,0) chirality does not contribute to the photocurrent in our sample, this is irrelevant given the large number of chiralities in a s–SWCNT sample with diameter of 0.75–1.2 nm. For large diameter nanotubes with $d > 1.2$ nm, the LUMO levels of PCBM and s-SWCNT are nearly degenerate, and the LUMO orbital is seen to extend across the interface (Figure 2a), thus yielding type-I alignment. We deduce that a maximum $d \approx 1.2$ nm needs to be employed for favorable PV operation in a s–SWCNT / PCBM active layer. Based on these trends, a rectifying behavior with non–zero PV efficiency is expected for a BHJ device employing PCBM and s–SWCNT with $d \approx 0.75$–1.2 nm in the active layer, while ohmic behavior and no PV effect are expected in a similar device employing s–SWCNT with $d > 1.2$ nm.

We have verified experimentally this prediction by preparing BHJ solar cells using PCBM and diameter–sorted s–SWCNT with both diameter ranges, and found excellent agreement with the interface type predicted using DFT, both for $PC_{60}BM$ and $PC_{70}BM$ (Figure 2b-c). In this first part of our study, we thus conclude that large diameter s–SWCNT in combination with $PC_{60}BM$ cannot provide PV conversion, while a suitable carbon–based solar cell can be prepared using a two–phase active layer of PCBM and s–SWCNT with $d \approx 0.75$–1.2 nm. However, in the small diameter s-SWCNT / PCBM devices we fabricated, the high van der Waals attractive force between small–diameter nanotubes caused extensive nanotube bundling and poor overall morphology (Figure 3a), and we were only able to achieve a maximum efficiency of ~0.4% using either $PC_{60}BM$ or $PC_{70}BM$ in combination with small–diameter nanotubes (Table 1).

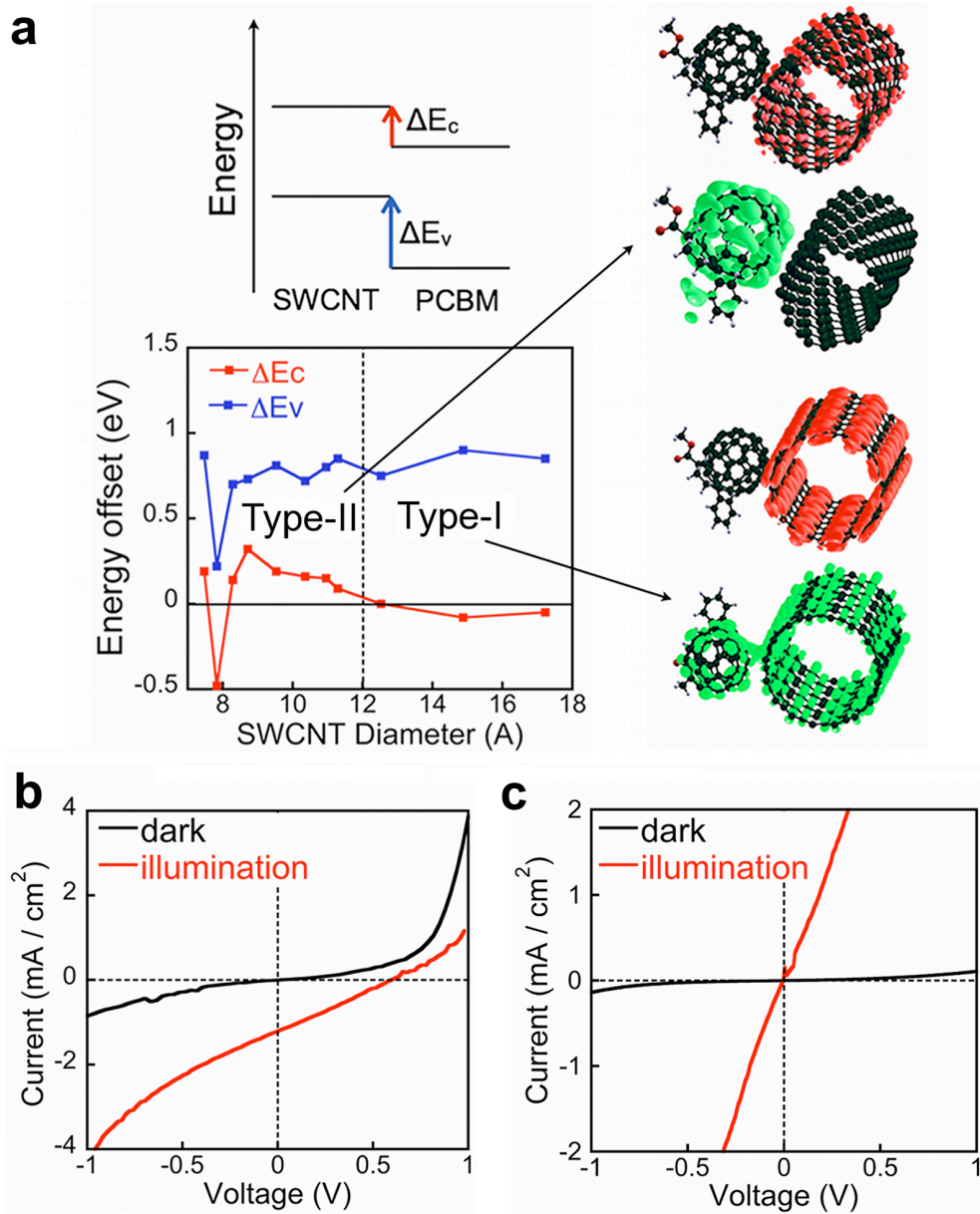

**Figure 2.** (a) Energy offsets of HOMO ($\Delta E_V$) and LUMO ($\Delta E_C$) levels as a function of s–SWCNT diameter calculated using DFT. The offsets are referenced to the HOMO and LUMO levels of the acceptor, as shown above the plot. The two diameter ranges with type–I and type–II alignment are indicated and delimited by a vertical dashed line at $d$ = 1.2 nm. The smallest diameter shown is the (6,5) nanotube, with type-II alignment and $\Delta E_C \approx 0.2$ eV. Also shown are charge density plots for the HOMO and LUMO orbitals for type-II and type-I cases, as indicated by the arrows. The band offset predictions are confirmed by current-voltage characteristics, shown in (b) for devices with nanotube diameter of 0.75–1.2 nm, and in (c) for nanotubes with $d$ > 1.2 nm. The same trends are found if $PC_{70}BM$ is used instead of $PC_{60}BM$ (see Table 1).

We obtained a much finer and smoother active layer morphology for nanotubes of larger diameter $d > 1.2$ nm blended with PCBM (Figure 3b); however as mentioned before these blends are unsuitable for PV operation due to type–I alignment. In order to take advantage of the favorable morphology of this blend, we added rGO as a third material phase that can induce exciton dissociation at the interface with PCBM or s–SWCNT. We found that this three-phase combination of PCBM / rGO / s–SWCNT with $d > 1.2$ nm can simultaneously achieve favorable morphology and exciton dissociation due to the formation of Schottky barriers at the interface with rGO as discussed below, and constitutes the active layer material of our best-efficiency device shown in Figure 1.

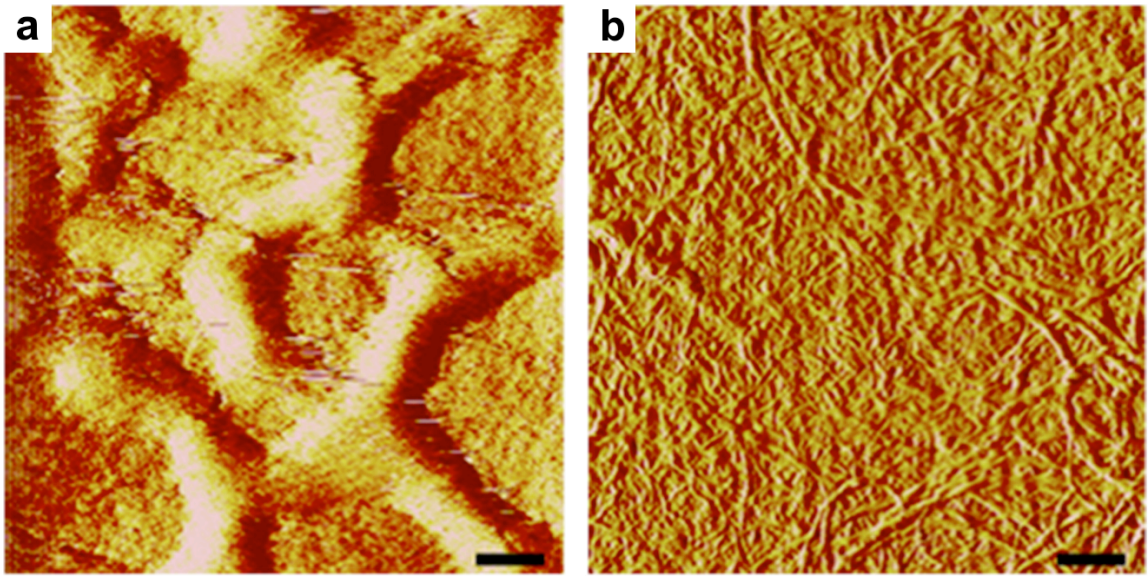

**Figure 3.** (a) AFM phase image of a device with active layer of PCBM / s–SWCNT with $d \approx$ 0.75–1.2 nm, showing the presence of nanotube bundles of up to 200 nm width. (b) Same image as in (a) but for a device made with larger diameter nanotubes ($d > 1.2$ nm), where the active layer components form a well-blended morphology without the formation of large nanotube bundles. Similar morphologies are found upon addition of rGO in both cases. The scale bar is 200 nm in both images.

In order to understand the role of rGO in enabling exciton dissociation in the three-phase system shown in Figure 1, we computed the electronic structure of rGO sheets with disordered oxygen–containing groups and O concentrations in the range of 10–20 at. % (as employed here experimentally), as well as the electronic structure of rGO / s-SWCNT and rGO / PCBM interfaces. In all cases considered here, rGO is found to be overall metallic, though due to the presence of local energy gaps it should be regarded as a highly disordered amorphous semiconductor, consistent with previous experimental observations.[22] For the interfaces between rGO and PCBM or s–SWCNT, our calculations predict the formation of n-type Schottky barriers larger than 1 eV for electrons to be transferred from PCBM to rGO (Supplementary Figure S1), and low p-type barriers for holes to be transferred from PCBM to rGO. Thus the transfer of holes from PCBM (with ionization energy of ~6.0 eV) to rGO (for which we calculated a work function of 5.1–5.3 eV for O concentrations of 10 – 20 at. %) is both kinetically and energetically favorable, while the transfer of electrons photogenerated within PCBM is kinetically hindered by large Schottky barriers. We also found small (~0.2 eV) p-type Schottky barriers for holes to be transferred from rGO to s–SWCNT (with ionization energy of ~4.8 eV for $d > 1.2$ nm). Based on these calculations, an operating mechanism for the best–efficiency device is suggested in Figure 1d, whereby holes photogenerated in PCBM (responsible for most of the photocurrent, see Figure 1c) are selectively transferred to rGO due the large Schottky barrier for electrons at this interface, and then to s–SWCNT due to a low hole barrier. The energetic of this process is favorable due to decreasing hole energy along this path, and due to the small kinetic barriers encountered as predicted by our calculations. A similar carrier–transfer cascade mechanism involving rGO interfaces has been recently suggested in the literature in a different system.[28] We note that in contrast the addition of rGO has a negative impact on PCBM / s-SWCNT devices with small diameter nanotubes (Table 1 and Supplementary Figure S2), due to the poor active layer morphology and to the fact that exciton dissociation is already possible in these devices in the absence of rGO because of type-II alignment.

Next, we describe the photodegradation (PD) behavior of carbon–based solar cells prepared in this work. PD is an open technological problem in polymer–based solar cells, requiring tight encapsulation and leading to device failure. Recent studies have significantly contributed to the understanding of PD patterns in polymer–based solar cells, by showing the presence of a burn–in process leading to the rapid initial degradation of encapsulated devices.[29]

In Figure 4a we compare the PD of PCBM / s–SWCNT / rGO carbon–based solar cells with that of a polymer–based P3HT / PCBM device[4] prepared and tested at the same time, under natural illumination and exposed to ambient environment *without encapsulation*. The P3HT / PCBM device shows a burn–in process leading to a rapid efficiency drop in the first 100 h, while the carbon–based device shows a gradual efficiency decrease without burn–in, contributed in equal measure from a decrease in $J_{sc}$ and $V_{oc}$ (Supplementary Figure S3). After aging both devices for 500 h, the efficiency of the polymer–based device reduces to ~15% of the initial value *versus* a much lower reduction to ~50% of the initial value for the carbon–based solar cell. The fact that the burn–in process only appears in devices where the polymer is present is in agreement with the findings in ref. 29 that attribute the rapid initial efficiency decrease to the presence of the polymer.

If oxygen and moisture are eliminated by carrying out the same aging test in a nitrogen glove box (Figure 4a), the efficiency only decreases by 5–10% over 500 h for the carbon–based solar cell (both with and without rGO), while a decrease by as much as 50% is observed for the polymer–based device. The significantly lower PD rate for carbon–based devices in a nitrogen environment compared to air suggests that oxidative processes due to the presence of PCBM are responsible for the PD in air seen in Figure 4a, consistent with recent PD studies of PCBM.[30] The residual PD for the polymer–based solar cell in nitrogen suggests on the other hand that polymer PD is partially contributed by optical excitation processes independent of the presence of oxygen.

Finally, we show in Figure 4b an effect unique to carbon–based PV devices: upon thermal heating in vacuum for 10 min of a PCBM / s–SWCNT / rGO solar cell previously aged for 500 h in air, the efficiency can be restored to 85% of its initial value if the proper temperature range is chosen for the annealing process, likely due to the partial reversibility of the PD oxidative processes. The same effect is not observed for a polymer–based solar cell (Figure 4b).

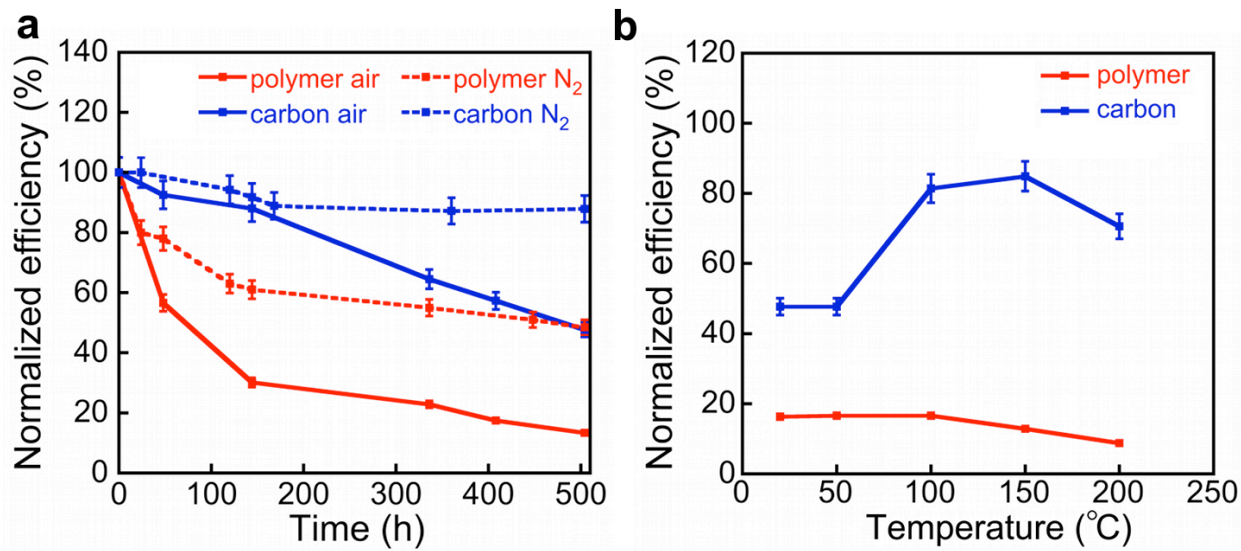

**Figure 4.** (a) Aging in air and in nitrogen of non-encapsulated solar cells, quantified by the time evolution of the percent fraction of the initial efficiency for the given device. A P3HT polymer / PCBM solar cell (referred to as "polymer" in figure) is compared with a SWCNT / PCBM / rGO solar cell ("carbon" curves in figure). A standard deviation of ~5% is shown for all the efficiency values reported here, based on a statistical sample of ~10 devices tested under the same conditions for each case. (b) Thermal annealing at different temperatures of devices previously aged in air for 500 h. An optimal annealing temperature of 150 °C allows to partially reverse the photodegradation of carbon–based devices, by restoring the efficiency up to 85% of the initial value. The same reversibility is not observed in a polymer–based device.

## DISCUSSION

There is significant potential to increase the efficiency of carbon–based solar cells beyond the 1.3% value reported here. The theoretical efficiency limit[31] of such devices depends on the absorption spectrum and energy gap of the materials present in the active layer, as well as on the maximum voltage achievable and on recombination. For the s–SWCNT / PCBM fabricated in this work, we estimate efficiency limits of 9% and 13%, respectively, for nanotube diameters of 1.2–1.7 nm and 0.75–1.2 nm, as derived in the analysis below.

The main active layer constituent (~90 wt. %) in our study is $PC_{60}BM$ (or $PC_{70}BM$), which contributes to most of the photocurrent and quantum efficiency as shown in Figure 1c. This also implies that absorption is limited to photons with energy above the optical gap of PCBM – apart from a small contribution from s-SWCNT absorption in the infrared, neglected here – and we can thus approximate the maximum short–circuit current $J_{sc}$ as:

$$J_{sc} = e \int_{E_{opt}}^{\infty} J_{ph}(E) \, dE$$

where $E_{opt}$ is the optical gap of PCBM (~1.8 eV), $J_{ph}$ is the incident solar photon flux for the AM 1.5 spectrum and $E$ is the incident photon energy. For the case of PCBM absorber, this yields an upper limit for $J_{sc}$ of 19.6 mA/cm$^2$ (see ref. 32) assuming 100% external quantum efficiency.

For an excitonic solar cell with exciton dissociation occurring at donor-acceptor heterojunctions, the maximum $V_{oc}$ can be approximated by the interface energy gap, namely the energy difference between the interface HOMO level (contributed here by the SWCNT donor) and the interface LUMO level from the PCBM acceptor. For small values of $\Delta E_C$ (e.g. ~0.2 eV as found here in Figure 1a), the maximum interface energy gap is approximately equal to the s–SWCNT energy gap, here taken as the average for the considered diameter range. This leads to limit $V_{oc}$ values of approximately 1.0 V for nanotube diameters of 0.75–1.2 nm and 0.7 V for nanotube diameters 1.2–1.7 nm, in agreement with the $V_{oc}$ found in this work. A practical limit for the fill factor (FF) can be taken as 0.65 as done in estimating limits for polymer solar cells;[33] this is not far from the

0.55 value obtained in our best device (Table 1). Finally, the limit efficiency $\eta_{max}$ under

AM1.5 illumination is calculated as the product:

$$\eta_{max} = \frac{J_{sc} \cdot V_{oc} \cdot FF}{P_{inc}}$$

where $P_{inc}$ is the incident power of 100 mW/cm$^2$ for AM1.5 illumination. Using the

values reported above, we obtain $\eta_{max}$ = 9% for nanotubes with large diameter of 1.2–1.7

nm, and $\eta_{max}$ = 13% for nanotubes with small diameter of 0.75–1.2 nm. We remark that

the alternative option of using a limit fill-factor value of $FF$ = 0.85 and a limit EQE =

0.85 (both of which have already been achieved in optimized organic solar cells, see ref.

34) would lead to limit efficiency values 10% higher than those reported here. In

comparing these limit values to the performance of the devices fabricated in this work,

we note that while $V_{oc}$ is near-optimal despite the limited efforts in our work to optimize this parameter, $J_{sc}$ is 7 – 10 times lower than the limit value of ~20 mA/cm$^2$. We believe that the main challenges towards significantly increasing $J_{sc}$ are the fabrication of thicker active layers than shown here (only ~100 nm) and the use of materials with higher purity to minimize recombination.

It must be noted that the limit efficiencies of 9 – 13% predicted here for

nanocarbon-based PV exceed those predicted for polymer-solar cells based on PCBM

acceptor, where $\eta_{max} \approx 11\%$ has been estimated and almost practically achieved after a decade of intense research. The main reason for this is that the optical gap of PCBM is similar to that of common polymer absorbers such as P3HT. The added benefit of using carbon as shown in this work is however the superior photostability, a highly desirable feature for solar cells with market potential. An optimization effort similar to the one for polymer-based solar cells should be undertaken for carbon-based devices as well, by identifying high-mobility and high purity semiconducting carbon materials – both bulk and nanostructured allotropes – amenable to low cost mass-fabrication.

# CONCLUSION

In summary, our results demonstrate that carbon–based PV active layers free of conjugated polymers or small molecules constitute a promising novel direction for photostable, efficient, solution–processable, thin–film, solar cells that are amenable to large–scale manufacturing. Candidate active layer materials are not limited to nanotubes and fullerenes as shown here, but rather span a vast array of suitable carbon compounds with yet untapped potential for thin–film solar cells. In combination with recently reported carbon–based transparent electrodes, carbon PV active layers could enable in the near future the development of efficient "all–carbon" solar cells.

# METHODS

**Preparation and Characterization of Carbon–Based Solar Cell Devices.** We used two distinct samples of high purity s–SWCNT (IsoNanotube–S from NanoIntegris, 98% purity) with different diameters in the range of, respectively, 0.75–1.2 nm and 1.2–1.7 nm and lengths in the 300 nm – 5 µm range in both cases. The nanotube sample with $d =$ 0.75–1.2 nm contained up to 50% of (6,5) nanotubes. Fullerene derivatives (6,6)-phenyl-C(X+1)-butyric acid methyl ester (PC$_X$BM), with X=60 or 70 were purchased from Sigma Aldrich. Graphene oxide was prepared in–house by a modified Hummers method starting from graphite powder (Bay Carbon, SP–1). The as-synthesized GO was dispersed in dimethylformamide and sonicated for 20 min, and then reduced by heating in an oil bath at 150 °C for 1 h. The oxygen concentration of rGO employed for device fabrication was determined to be in the range of ~10–20 at. % by Fourier transform infrared (FT–IR) and NMR spectroscopies. Carbon–based PV devices were prepared in a nitrogen glove box and consisted of the following sequence of films and thicknesses: ITO/PEDOT:PSS(40 nm)/TFB (10 nm)/CPV (120 nm)/Al (100 nm), where 1) ITO is Indium Tin oxide (20 Ohm/Sq from Thin Film Devices), in the form of a 0.5 x 0.5 in$^2$ glass substrate with pre-patterned ITO electrodes; 2) PEDOT:PSS is Poly(3,4-ethylenedioxythiophene) poly(styrenesulfonate) (Clevios), a transparent hole conducting layer spun cast onto ITO; 3) TFB is poly[(9,9-dioctylfluorenyl-2,7-diyl)-co-(4,4'-(N-(4-sec butylphenyl)) diphenylamine)] (American Dye Source), an electron blocking layer; 4) CPV is a blend of carbon nanomaterials including s–SWCNT, PCBM and rGO with different compositions as shown in Table 1. Carbon nanomaterials were dissolved, sonicated and filtered in 1,2-dichlorobenzene, spun cast at 600 rpm for 60 s, and allowed to solvent anneal overnight. The sonication was performed using a low-power bench-top sonicator for 30 min without heating. Such mild sonication conditions do not lead to formation of the ODCB polymer, as confirmed by high-resolution transmission electron microscopy. 5) Al metal was evaporated at a rate of 1 nm/s as the top contact layer. The final device area was defined by the overlap between the top and bottom electrodes. Current–voltage (I–V) characteristics of the devices were measured in a glove box with a source-meter (Keithley 6487). The PV performance (power conversion efficiency and $I$–$V$ curves) was measured under illumination from a 100 mW/cm$^2$, AM1.5 solar simulator. Over 10 devices of each kind reported in Table 1 were fabricated and tested, yielding a 5% standard deviation on the efficiency. Transmittance and absorbance spectra of the device active layer were measured with a Shimadzu UV-Vis-NIR dual-beam spectrophotometer (UV-3600). External quantum efficiency (EQE) measurements were collected in a glove box using a monochromator, chopped locked-in, and an NREL calibrated Ge detector (Newport). Surface morphology of the devices was investigated using an atomic force microscope (Digital Instruments Dimension 3000).

**Computational Design of Carbon Interfaces**. Density functional theory calculations were performed using the VASP[35] and Quantum Espresso codes[36] with the Perdew-Burke-Ernzerhof exchange-correlation functional.[37] A kinetic energy cutoff of 35 Ry was used for the plane-wave basis set and of 200 Ry for the charge density, and all structures were relaxed to less than 50 meV/A in their residual atomic forces. For the s-SWCNT / PC$_{60}$BM interfaces, ultrasoft psuedopotentials[38] were used to describe the core electrons. An orthorhombic unit cell with 10 A vacuum in the non-periodic directions was employed. Between 1–4 SWCNT repeat units (depending on the nanotube chirality) were used with a Monkhorst-Pack $k$-grid of 1 x 1x $n_z$ , where $n_z$ values up to 20 were used for a converged number of $k$-points in the nanotube axis direction. The PCBM was placed at a van der Waals distance of ~3.3 A next to the nanotube and relaxed to eliminate residual forces; the final calculations were carried out on such combined PCBM / s–SWCNT systems with up to ~500 valence electrons. The HOMO and LUMO level offsets were computed as differences in the peaks of the projected density of states obtained for the two molecules in the combined system. For electronic structure and workfunction calculations on rGO, and for the determination of Schottky barriers (SB) at rGO / PCBM and rGO / s–SWCNT interfaces, projector augmented–wave (PAW) pseudopotentials[39] were employed as implemented in the VASP code.[35] Representative rGO structures were assumed to consist of epoxy, hydroxyl, carbonyl and ether groups, and were generated by placing functional groups in different proportions on a square cut of the graphene sheet (180 C atoms) by using a random number generator, similar to the method used in ref. 40. The structures were first relaxed using the MMFF94 force field and further relaxed in DFT. The oxygen content was varied between 10–20 at. % to consider structures with the same O concentration range as in the experiment. For the rGO / PCBM and rGO / s–SWCNT interface calculations, the PCBM (or s–SWCNT) was placed at a van der Waals distance of 3.5–4.0 A from the graphene basal plane depending on the O content of the rGO structure. Only the p-type Schottky barrier ($E_p$) at the interface was calculated directly using DFT with a method detailed in ref. 41. The n-type Schottky barrier ($E_n$) in the case of rGO / PCBM was obtained using the relation, $E_n = E_g - E_p$, where $E_g$ is the experimental electronic gap of PCBM (~2 eV, see ref. 27), and where we used the fact that the sum of the n-type and p-type barriers is numerically equal to the semiconductor energy gap. No assumptions were made about the position of the Fermi energy, and chemical doping was not assumed. For each concentration, the SB was obtained as the average value over three different PCBM (or s–SWCNT) positions in the simulation cell, with a small standard deviation of ~0.1 eV.

*Conflict of Interest:* The authors declare no competing financial interests.

*Acknowledgement:* M.B. acknowledges funding from Intel through the Intel Ph.D. Fellowship. M.B., N.F. and J.C.G. thank ENI for funding through the MITEI program. M.B., P.V.K. and J.C.G. wish to thank NERSC and Teragrid for providing computational resources. S.R. thanks the University of Kansas for its startup financial support, and acknowledges funding from a Department of Energy award (DESC0005448).

*Author Contributions:* M.B. and J.C.G. developed the concept. M.B. designed and performed calculations, contributed to designing the experiments, analyzed the data, and wrote the manuscript. J.L., A.K. and S.R. designed and performed experiments, and analyzed the data. S.R. and J.C.G. contributed to writing the manuscript. P.V.K. performed calculations involving rGO. N.F. contributed to the discussion and photodegradation experiments.

*Supporting Information Available:* Computed Schottky barriers. EQE for small diameter nanotube solar cells. Additional photodegradation data. This material is available free of charge *via* the Internet at http://pubs.acs.org

calculations with non-local functionals (*e.g.* hybrid DFT or GW self-energy methods) are however unfeasible at present. Nonetheless the error cancellation leads to qualitatively correct trends and band offsets predictions as verified experimentally in Figure 2.